\documentclass[%
prb,
 reprint,
 amsmath,amssymb,superscriptaddress
]{revtex4-1}
\usepackage{xparse}

\usepackage{appendix}
\usepackage{graphicx}
\usepackage{dcolumn}
\usepackage{bm}

\newcommand{\ket}[1]{\ensuremath{\left| #1 \right>}}
\newcommand{\kety}[1]{\ensuremath{\left| #1 \right)}}
\newcommand{\bra}[1]{\ensuremath{\left< #1 \right|}}
\newcommand{\bray}[1]{\ensuremath{\left( #1 \right|}}
\newcommand{\braket}[2]{\ensuremath{\left< #1 \ \vphantom{#2} \right| 
\left. #2 \vphantom{#1} \right>}}

\newcommand{\matrixel}[3]{\ensuremath{\left< #1 \vphantom{#3} \right| #2 
\left| #3 \vphantom{#1} \right>}}

\usepackage{bbold}

\begin{document}

\title{Quantum Work of an Optical Lattice}

\author{Colin Rylands}
\email{crylands@umd.edu}

\affiliation{Joint Quantum Institute and Condensed Matter Theory Center, Department of Physics,
University of Maryland, College Park, Maryland 20742-4111, U.S.A.}
\author{Natan Andrei}
\email{natan@physics.rutgers.edu}
\affiliation{Department of Physics, Rutgers University,
Piscataway, New Jersey 08854, U.S.A.
}

\begin{abstract}
 A classic example of a quantum quench concerns the release of a interacting Bose gas from an optical lattice. The local properties of quenches such as this have been extensively studied however the global properties of these non-equilibrium quantum systems have received far less attention. Here we study  several aspects of global non-equilibrium behavior by calculating the amount of work done by the quench as measured through the work distribution function. Using Bethe Ansatz techniques we determine the Loschmidt amplitude and work distribution function of the Lieb-Liniger gas after it is released from an optical lattice. 
We find the average work and its universal edge exponents from which we determine the long time decay of the Loshcmidt echo and highlight striking differences caused by the the interactions as well as changes in  the geometry of the system. We extend our calculation to the attractive regime of the model and show that the system exhibits properties similar to the super Tonks-Girardaeu gas. Finally we examine the prominent role played by bound states in the work distribution and show that, with low probability, they allow for work to be extracted from the quench.

\end{abstract}

\maketitle
\section{Introduction}
The quantum quench is one of the simplest protocols of non-equilibrium quantum physics.  An adiabatically closed system is initially  prepared  in some  state $\ket{\Psi_i}$, typically an eigenstate of a Hamiltonian $H_i$. At a given moment, $t=0$, the parameters of the system are suddenly changed and the system  time evolves under a new Hamiltonian $H$. The sudden quench of the system parameters from  the initial to  the final Hamiltonian  excites states throughout the spectrum and in doing so a truly non equilibrium situation is created\cite{CalabreseCardy1, PolRev, MitraRev}.  A classic quantum quench experiment concerns the release of gas or a Bose Einstein condensate from an optical lattice which is suddenly removed allowing the gas to expand  \cite{IBRMP, BlochCollapse,WillExp}. Experiments of this type address many questions that concern the dynamics of these systems, their   entanglement, entropy production or thermalization, to name a few.

Low dimensional systems are of particular interest in this regard. The enhanced quantum fluctuations in such systems give access to the study of various strongly correlated phases that are  hard to reach in higher dimensions. Many of these low dimensional systems are described by integrable Hamiltonians which facilitates their study  by powerful analytic methods such as the Bethe Ansatz and conformal field theory \cite{AndreiQuench, VidmarRev, CalabreseCardyJstat, EsslerFagottiJstat, CauxJstat}. A  major  focus  has  been  the study of the  local  properties of the quenched system and in particular  the behavior of local observables and correlation functions.
In parallel, it was understood that a quench constitutes a thermodynamic process and within this context one can examine  the concepts of work, entropy and heat of a far from equilibrium quantum system \cite{TakLutHan, Silva08,Goold, QHE, DecampQHE}. These  global properties of post quench systems will be our main concern here. In particular,  we will consider a  gas of neutral bosonic atoms in a one dimensional trap described by the  Lieb-Liniger Hamiltonian. The atoms are initially in the ground state of a 1D optical lattice and then suddenly released. We will calculate the work distribution of the  quench, examining both the repulsive and attractive regimes as well as open and periodic boundary conditions.

As always,  the work is given by the difference between two measurements of the energy, one pre- and the other post- quench, $W=E_f-\epsilon_i$. \footnote{$dW= d E$, with $dQ=0$ as the system is isolated.} But while the initial energy $\epsilon_i$ is given, the final energy may be any of the eigenvalues of the post quench Hamiltonian, $E_n$ which can be measured with probability, $P_n=|\, \langle n|\Psi_i\rangle \,|^2$. This renders quantum work a random variable with a probability distribution  defined as, \cite{TakLutHan, Silva08} 
\begin{eqnarray}\label{Pdef}
\mathcal{P}(W)=\sum_{n} \delta\left(W-(E_n-\epsilon_\text{i})\right)|\braket{n}{\Psi_i}|^2.
\end{eqnarray}
Here $\ket{n}$ are  the eigenstates of $H$ with energy $E_n$ and $\epsilon_\text{i}$ is the initial energy.

Much is known about the form of this distribution function in a number of models and quenches\cite{Silva08, GamSil1, GamSil2, SotGamSil,Smacchia, delCamp} including certain limits and approximations of the optical lattice quench discussed above\cite{Sotiriadis, Palmai,Palmai2,RylandsMTM}. In the  study of ref \citenum{RylandsMTM}, the optical lattice is lowered but not removed entirely unlike the case we will consider here. The retention of the lattice is a crucial component  as then the final Hamiltonian remains gapped and there exists a basis of long lived quasi-particles. These two features allow for a very intuitive picture of the work distribution to emerge. 
$\mathcal{P}(W)$ is defined for $W\geq \delta E$ with $\delta E= E_0-\epsilon_\text{i}$ being  the  energy  difference  between  the  ground state of $H$  and the initial  state. It possesses a delta function peak at $W=\delta E$  weighted by the fidelity, $\mathcal{F}=|\braket{\Psi_i}{0}|^2$ signifying a transition from the initial state to the ground state. Separated from this there  exists a continuum of excited states into which $\ket{\Psi_i}$ can transition during the quench. The lower threshold for the  continuum is at $W=2m+\delta E$ with $m$ being the mass of the lightest quasi-particle. This signifies the emission of two quasiparticles with opposite momentum from the initial state.  At the threshold, $\mathcal{P}(W)$  exhibits an edge singularity similar to the Anderson and Mahan effects in the X-ray edge problem\cite{Mahan}. When quenching from the ground state of the optical lattice in a non interacting model the  distribution diverges at the threshold with exponent $\alpha=-1/2$\cite{SotGamSil}. On the other hand when interactions are present  (even weak ones)  this changes drastically to a square root singularity, $\alpha=1/2$, exemplifying the strongly correlated nature of interacting one dimensional systems \cite{ Palmai2, RylandsMTM}. Above this, similar edge singularities will occur when new excitation channels open up e.g. the emission of 2n particles with zero momentum causes an edge singularity at $W=2nm+\delta E$ with an exponent $n/2$ when interactions are present.
Below the threshold, in the region $\delta E<W <2m+\delta E$ additional delta function peaks appear if the theory  supports bound states. These occur at $W=m_b+\delta E$, $m_b$ being the masses of the bound states which have the same parity as the initial state. For the ground state quench only bound states comprised of an even number of particles will appear\cite{RylandsMTM}. In the thermodynamic limit the distribution becomes peaked about the average work with fluctuations vanishing as $1/\sqrt{N}$ with the most interesting features found in the region of the threshold singularity \cite{SotGamSil}.

\begin{figure}
 \includegraphics[width=.35\textwidth]{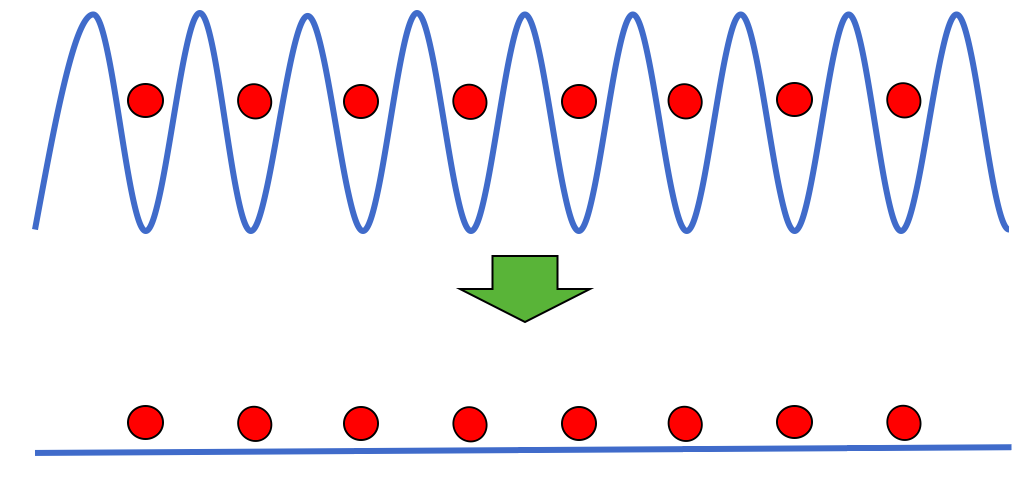}
\caption{The optical lattice quench. An interacting Bose gas is held in a deep optical lattice with at most a single boson per site. The lattice is then suddenly removed and the gas allowed to expand. We calculate probability distribution of the work done on the gas during this process. }\label{FigSchematic}
\end{figure}

Here, in contrast to study described above , we will examine the work done when the lattice is removed entirely with the post quench evolution  governed by the Lieb-Liniger model, see figure \ref{FigSchematic}. Since this model is gapless, the perspective of the work distribution function just elucidated is no longer correct. In a gapless theory the quench creates a macroscopic number of excitations making studies of the work statistics much more difficult\cite{VasseurRL, PerfettoPiroliGambassi} prompting some natural questions. What is the fate of the edge singularities when the gap is reduced to zero? Do the exponents change  and and is the dependence on interactions still so dramatic? How do bound states present themselves if the model is gapless, do they just melt into the continuum? In what follows we  will examine these questions as well as discuss the prominent role played by boundary conditions in this quench.

The remainder of the paper is organized as follows. In section II we introduce the model and our quench protocol. A useful identity is presented and used to calculate the time evolution of the initial state. In section III the Loschmidt amplitude is calculated for arbitrary coupling strength and particle number. From this we determine the work distribution function and examine it for strong repulsive interactions, finding the average work as well as the universal edge exponents. This analysis is extended to the attractive regime where we examine how bound states change the distribution and allow for negative values of the work to be measured. In the penultimate section we contrast the behavior in systems where periodic boundary and open boundary conditions are imposed. Finally we summarize our work, discuss generalizations of the results as well as relevance to experiment.

\section{An alternate identity}
An excellent description of a cold atomic gas in the absence of an external potential is furnished by the Lieb-Liniger (LL) model\cite{IBRMP}. The Hamiltonian is
\begin{eqnarray}\nonumber\label{HLL}
H=\int\mathrm{d} x\, \Bigg\{b^\dag(x)\left[-\frac{\partial_x^2}{2m}\right]b(x)+ c\,b^\dag(x)b(x)b^\dag(x)b(x) \Bigg\}
\end{eqnarray}
where $b^\dag(x),b(x)$ are creation and annihilation fields of  bosons which have a point-like density-density interaction of strength $c$ and we set $\hbar=1$. In this article we consider both the repulsive, $c>0$, and attractive, $c<0$, regimes. The model is well known to be integrable for all couplings $c$ \cite{LiebLin1, LiebLin2} and its (unnormalised) $N$-particle eigenstates can be expressed in the form
\begin{eqnarray}\label{states}
\int\mathrm{d}^Nx \prod_{i<j}\frac{k_i-k_j-i c\,\text{sgn}(x_i-x_j)}{k_i-k_j-i c}\prod_{l}^Ne^{ik_lx_l}b^\dag(x_l)\ket{0}.
\end{eqnarray}
The energy and momentum of such a state is $E=\sum_j^Nk_j^2/(2m),~P=\sum_{j=1}^Nk_j$ and if periodic boundary conditions are imposed the single particle momenta are quantized according to the Bethe equations,
\begin{eqnarray}\label{BAE}
k_j=\frac{2\pi}{L}n_j-\frac{1}{L}\sum_j^N\varphi(k_j-k_k).
\end{eqnarray}
Here $\varphi(x)=2\arctan{(x/c)}$ is the two particle phase shift, $L$ is the system size and $n_j$ are distinct integers or half integers which serve as the quantum numbers  labeling the eigenstates of the periodic system,  and unless otherwise stated it should be understood  that $k_j$ is the single particle momentum corresponding to $n_j$ according to \eqref{BAE}.

Our quench protocol consists of releasing the system from an initially deep optical trap. This deep trapping potential is modelled by taking the initial state to be 
\begin{eqnarray}\label{PSI0}
\ket{\Psi_i}=\int\mathrm{d}^Nx\prod_{j=1}^N \left[\frac{m\omega}{\pi}\right]^{\frac{1}{4}}e^{-\frac{m\omega}{2}(x_j-\bar{x}_j)^2}b^\dag(x_j)\ket{0}
\end{eqnarray}
which is the ground state of an optical lattice of frequency $\omega$ lattice spacing $\delta$. The bosons are initially taken to be located at positions $\bar{x}_j$ with $\bar{x}_j-\bar{x}_{j+l}=\delta$. We restrict to the situation where there is at most one boson per site with the sites  filled consecutively thus allowing for any value of $\rho=N/L$  provided $\rho\leq \delta$.  It is further assumed that the trap is deep enough that any overlap between neighbouring sites is negligible. The initial state is then evolved according to the LL Hamiltonian.

Standard practice is to study the evolution of the initial state, $e^{-iH_{\rm LL}t}|\Psi_i \rangle$, by inserting a resolution of the identity in the basis of the many-body system eigenstates,
\begin{eqnarray}\label{Identity}
\mathbb{1}_N=\sum_{n_1<\dots<n_N}\frac{\ket{\{n\}}\bra{\{n\}}}{\mathcal{N}(\{n\})}
\end{eqnarray}
 with $\mathcal{N}(\{n\})$ being the norm of the Bethe states. 
 For repulsive interactions this  is given by the Gaudin formula\cite{Gaudin, Korepin}
\begin{eqnarray}\nonumber
\mathcal{N}(\{n\})=\det\left[\delta_{jk}\left(L+\sum_{l=1}^N\varphi'(k_j-k_l)\right)\!-\!\varphi'(k_j-k_k)\right].
\end{eqnarray}
After
calculating the overlaps $C_n= \langle\{n\}| \Psi_i\rangle/\mathcal{N}$ the  time evolution can  be trivially performed according to  $\ket{\{n\},t}=e^{-iE(\{n\})t}\ket{\{n\}}$. The bottleneck in this procedure occurs in the calculation of the overlaps which proves to be rather difficult. Outside of the Tonks-Girardeau (TG) limit  of $c\to\infty$, where calculations are  simplified \cite{Pezer, vandenBerg, DynamicalDepin, DarkSolitonsTG, EquiibrationinTG,YukalovGirardeau}, exact overlaps are scarce \cite{Brockmann, deNardis}. To simplify the calculation we shall use an alternate resolution of the identity  that is particularly convenient when working with initial states like \eqref{PSI0} which are ordered in real space\cite{GoldsteinAndrei} , 
\begin{eqnarray}\label{IdentityY}
\mathbb{1}_N=\sum_{n_1,\dots,n_N}\frac{\ket{\{n\}}\bray{\{n\}}}{\mathcal{N}(\{n\}}.
\end{eqnarray}
Here we have  introduced the notation $\kety{\{n\}}$ to describe an eigenstate of \eqref{HLL} restricted to a certain ordering in real space,
\begin{eqnarray}\label{Yud}
\kety{\{n\}}=\int\mathrm{d}^Nx\,\theta(\vec{x})\prod_l^N e^{ik_lx_l}\ket{0}
\end{eqnarray}
where $\theta(\vec{x})$ is a Heaviside function which is non zero only for $x_1>x_2>\dots x_N$, and the  momenta $k_j$ are determined by the quantum numbers $\{n\}$. It is important to note also that the ordering in the  sum over quantum numbers of the system present in \eqref{Identity} has been removed in \eqref{IdentityY}. This alternate resolution of the identity is implicitly ordered in real space as opposed to momentum space and therefore is the natural choice for calculating overlaps such as those with $\ket{\Psi_i}$. Using the properties of the Bethe states\cite{Korepin} it can be  confirmed that this expression satisfies all the properties of a resolution of the identity.

The overlaps between \eqref{Yud} and \eqref{PSI0} are now straightforwardly calculated, allowing us to express the initial state as \begin{eqnarray}\label{PSIt}
\ket{\Psi_i}=\left[\frac{4\pi}{m\omega}\right]^{\frac{N}{4}}\sum_{n_1,\dots,n_N}\frac{e^{-\sum_{j=1}^N\left[\frac{k_j^2}{2m\omega}+ik_j\bar{x}_j\right]}}{\mathcal{N}(\{n\})}\ket{\{n\}}.
\end{eqnarray}
Before proceeding further we make some comments on the above expression. The apparent ease with which we have arrived at \eqref{PSIt} was facilitated entirely by the correct choice of identity and was further simplified by the fact that there was at most a single boson per site. 

The central goal of this paper is to study the amount of work done, $W$, when the optical lattice is lowered. More precisely we will calculate the work probability distribution \cite{Silva08, TakLutHan}. In the notation of \eqref{Identity} this is
\begin{eqnarray}
\mathcal{P}(W)=\!\!\sum_{n_1<\dots <n_N} \!\!\delta\left(W-(E(\{n\})-\epsilon_\text{i})\right)\frac{|\braket{\{n\}}{\Psi_i}|^2}{\mathcal{N}(\{n\})}
\end{eqnarray}
 In the present circumstances $\epsilon_\text{i}=N\omega/2$ and from here on we measure the work done staring from this value, $W\to W-\epsilon_\text{i}$.  
To proceed, we introduce the Loschmidt amplitude (LA) $\mathcal{G}(t)=\matrixel{\Psi_i}{e^{-iH t}}{\Psi_i}$ which is the Fourier transfrom of the work distribution
\begin{eqnarray}\label{p}
\mathcal{P}(W)=\int_{-\infty}^\infty\frac{\mathrm{d}t}{2\pi} e^{iWt}\mathcal{G}(t).
\end{eqnarray}
The LA is a quantity of significance in its own right and  central to a number of fields. The zeros of the LA define dynamical quantum phase transitions \cite{HeylKehrein, Heylreview}  whilst the square of the LA, $|\mathcal{G}(t)|^2$, alternately known as the Loschmidt echo or return rate is prominent in studies of quantum chaotic systems\cite{delCamp,LosChaos}. Predominantly, we shall employ it as a calculational tool to determine $\mathcal{P}(W)$.

Using the expression \eqref{PSIt} we find that 
\begin{eqnarray}\label{Gexact}
\mathcal{G}(t)&=&
\left[\frac{4\pi}{m\omega}\right]^{\frac{N}{2}}\!\!\sum_{n_1,\dots,n_N}e^{-\frac{1}{m\omega}\left[1+i\frac{\omega }{2}t\right]\sum_{j=1}^Nk_j^2}\frac{G(\{n\})}{\mathcal{N}(\{n\})}
\end{eqnarray}
where $G(\{n\})=\det{\left[e^{-ik_j(\bar{x}_j-\bar{x}_k)-i\theta(j-k)\varphi(k_j-k_k)}\right]}$ and $\theta(j-k)$ is a Heaviside function.
Written out explicitly this is
\begin{eqnarray}\nonumber
G(\{n\})=\sum_{P\in S_N}(-1)^Pe^{-i\sum_j^Nk_j(\bar{x}_j-\bar{x}_{Pj})-i\sum_{(j,k)\in P}\varphi(k_j-k_k)}
\end{eqnarray}
where the sum, $\sum_{P\in S_N}$, 
is over elements of the symmetric group and $(j,k)\in P$ is shorthand for pairs whose relative position is exchanged by the permutation, $j<k,~P(j)>P(k)$. 
This sum over permutations can be given the interpretation of particles exchanging positions after expanding from their original lattice positions with every exchange of particles being accompanied by the two particle phase shift, $\varphi$. 

The formula \eqref{Gexact} gives the exact Loschmidt amplitude for arbitrary $c, N$ and $L$ however its generality makes it somewhat cumbersome. One can simplify it by expanding in $c\gg m\omega$. Under this assumption $\varphi(x)\approx2x/c +\mathcal{O}(1/c^3)$ and we find that 
\begin{eqnarray}\label{BAEc}
k_j=\left[1+\frac{2\rho}{c}\right]^{-1}\frac{2\pi}{L}n_j,\\\label{Nc}
\mathcal{N}(\{n\})=L^N\left(1+(N-1)\frac{2\rho}{c}\right)
\end{eqnarray}
and  in \eqref{BAEc} we have used the fact that only states with zero momentum are present in the sum \eqref{Gexact}. For a finite size system the Loschmidt amplitude displays recurrences with a period $\tau= \left(1+2\rho/c\right)^2L^2/\pi\omega$. These recurrences disappear in the infinite volume limit.


\begin{figure}
 \includegraphics[width=.45\textwidth]{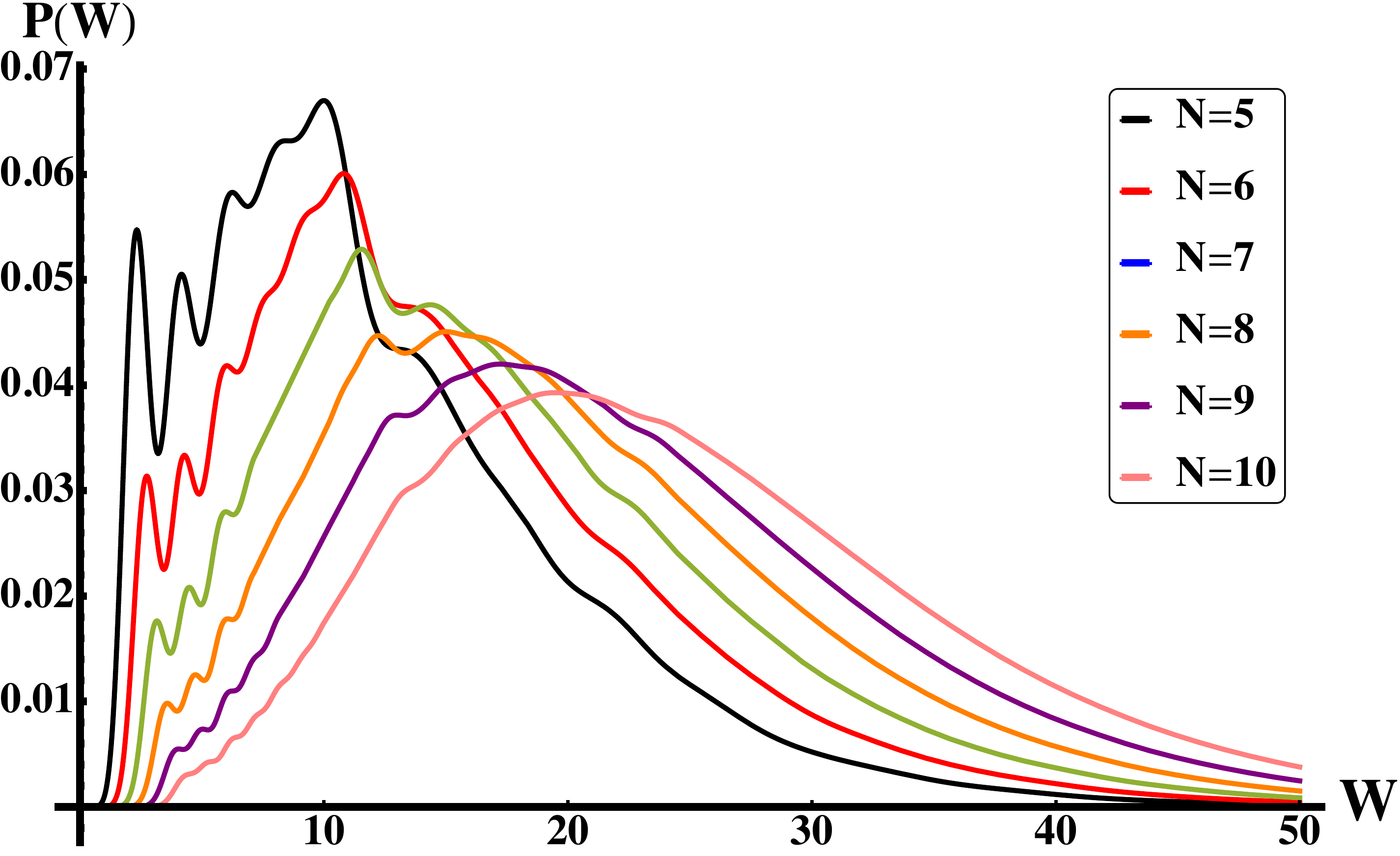}
\caption{The work distribution function, $\mathcal{P}(W)$, for particle number $5\leq N\leq10$ with $\delta/m=2$ and $\omega=10$ and repulsive interactions. The resonances at lower values of $W$ are washed out as the particle number is increased. Their position and size depends upon the interaction strength. }\label{Fig1}
\end{figure}

\section{Work in infinite volume}
We turn now to the evaluation of the work done by Fourier transforming the Loschmidt amplitude of the open system. This is the case where the occupied part of the lattice is much smaller than its overall size $\rho\ll\delta$. We shall also consider in the next section the  case of a fully filled lattice, $\rho=\delta$ and periodic boundary conditions. 

\subsection{Repulsive interactions}

In the thermodynamic limit, $N,L\to \infty$ the sum over quantum numbers becomes a product of integrals which can be evaluated giving,
\begin{eqnarray}\label{GQ}
\mathcal{G}(t)&=&
\frac{1}{\left[1+i\frac{\omega}{2}t\right]^{\frac{N}{2}}}\sum_{P}(-1)^Pe^{-\frac{\omega \alpha_P^2}{4\left(1+i\frac{\omega}{2}t\right)}}.
\end{eqnarray}
 Where we have introduced $\alpha_P^2=m\delta_\text{eff}^2\|P\|^2/2$ with $\delta_\text{eff}=\left[1+\frac{2}{c\delta}\right]\delta$, an effective distance between lattices sites and $\|P\|^2=\sum_j^N(j-P(j))^2$. Again we can interpret the sum over permutations as the a sum over particles exchanging positions, with $\|P\|^2=2$ for example corresponding to a neighbouring pair exchanging positions while $\|P\|^2=8$ could be 4 nearest neighbour exchanges or 1 next nearest neighbour exchange.
\begin{figure}
 \includegraphics[width=.45\textwidth]{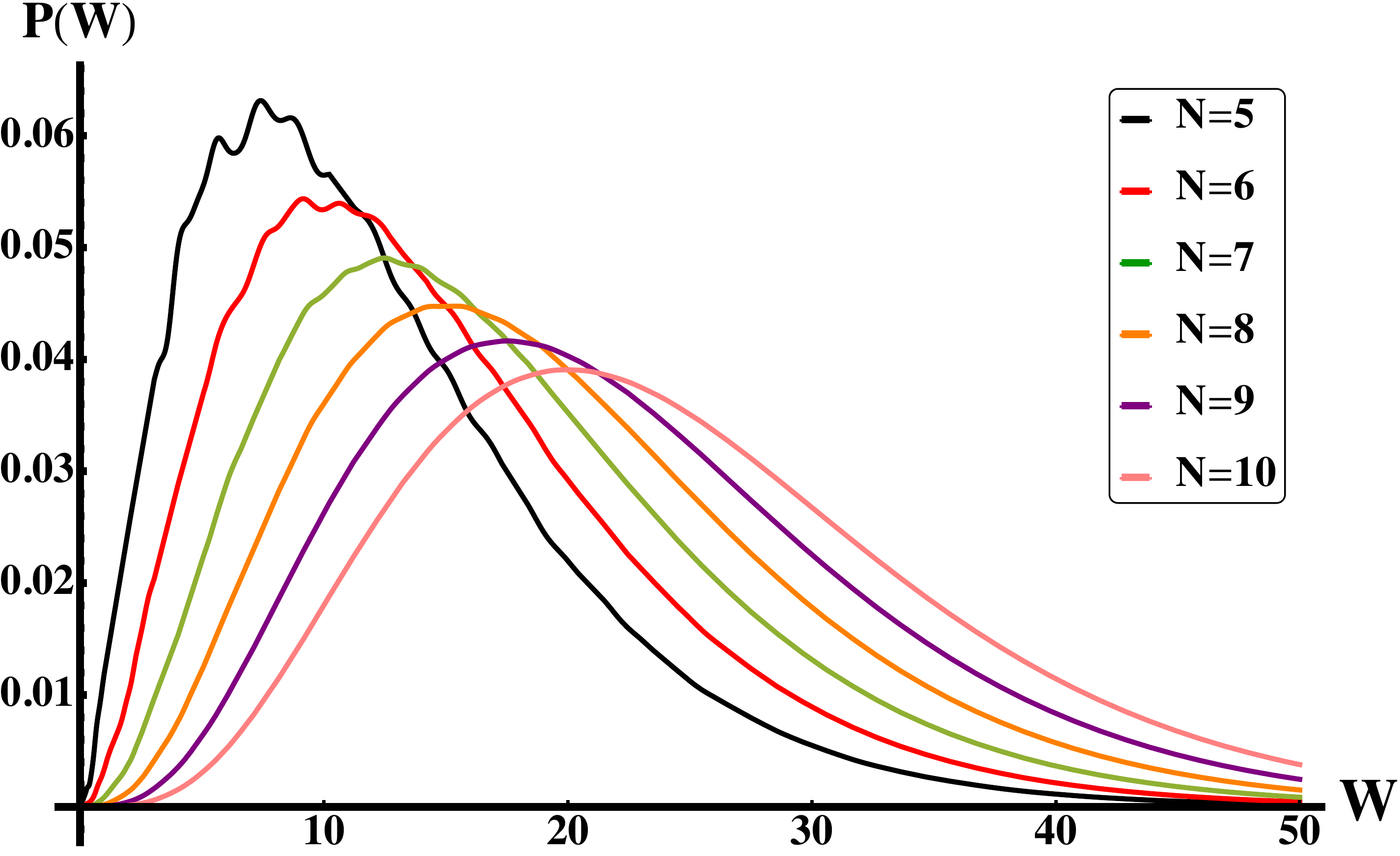}
\caption{The work distribution function, $\mathcal{P}(W)$, for particle number $5\leq N\leq10$ with $\delta/m=2$ and $\omega=10$ for free bosons. }\label{Fig2}
\end{figure} 
The large repulsive interaction will inhibit the spreading of the particles which is felt through an increase in the effective distance between sites.
Including higher order terms in this large $c$ expansion will cause further dressing of this distance. A similar dressing of the distance occurs in the scattering of soliton-like objects in integrable models\cite{QsolitonScattering, SolitonGases, Hardrods}.

 Performing the Fourier transform of $\mathcal{G}(t)$ we find that the work distribution function is
\begin{eqnarray}\label{Prep}
\mathcal{P}(W)&=&\frac{e^{-\frac{2W}{\omega}}}{W}\left[\frac{2W}{\omega}\right]^{\frac{N}{2}}
\sum_P(-1)^P\frac{J_{\frac{N-2}{2}}(2\sqrt{\alpha_P^2W})}{\left[\alpha_P^2W\right]^{\frac{N-2}{4}}}~~
\end{eqnarray}
where $J_n(x)$ is a Bessel function of the first kind. Here we see that the  sum over the $P$ is analogous to the sum over the number of excited particles for the gapped case which was discussed in the introduction. A notable distinction from the gapped case is the absence of a delta function peak as well as any threshold singularity at finite $W$. 

We plot \eqref{Prep} for different values of $N$ in Fig. \ref{Fig1} and also the non-interacting result for the same values in Fig. \ref{Fig2}. Comparing the two figures we see some common features as well as some striking distinctions. We note that the average of the distributions $\left<W\right>=\int \mathrm{d}W \,W\mathcal{P}(W)$  appears to be independent of the presence of interactions. Since  the quench is extensive in nature we can  expect that  $\left<W\right>\sim N$ which is seen in the figures through the rightward shift of the distributions. Using this along with the properties of the Bessel function we find  that the dominant contribution of the distribution in this region comes from the identity permutation
\begin{eqnarray}
\mathcal{P}(W)\sim\frac{e^{-\frac{2W}{\omega}}}{W\Gamma(N/2)}\left[\frac{2W}{\omega}\right]^{\frac{N}{2}}.
\end{eqnarray}
This is the moment generating function of the Gamma distribution whose average is  $\left<W\right>=N\omega/4$. As anticipated it is independent of the interaction strength.  The lack of dependence on $c$ can be understood from the latter formula along with the fact the initially the bosons have negligible overlap. This in agreement with the result calculated using $\left<W\right>=\matrixel{\Psi_i}{H}{\Psi_i}$.  Along similar lines one can show that the $m^\text{th}$ moments $\mathcal{P}(W)$, with $m\ll N$ are also independent of the interaction for example the variance and skewness are $N\omega^2/8 $ and  $ \sqrt{2/N}$ respectively. 
Going beyond this we can calculate the average exponentiated work\cite{Jarzynski1, TakLutHan}
\begin{eqnarray}
\left<e^{-\beta W}\right>=\left(\frac{1}{1+\frac{\omega\beta}{2}}\right)^{\frac{N}{2}}\left[1+\sum_{P\neq\mathbb{1}}(-1)^P\frac{e^{-\frac{\omega\alpha^2_P}{2+\omega\beta}}}{\Gamma(N/2)}\right]
\end{eqnarray}
from which we can derive all the moments of the work distribution through repeated differentiation of $\beta$.

The large $W\gg\left<W\right>$ regime appears to be also unaffected by the interactions which can be interpreted by translating back to the language of the the LA using \eqref{p}. At short times the bosons expand from their initial trap positions without encountering one another and so are independent of the interactions. 

At small values of $W$ the  distribution is strongly affected by the interactions. We can see large resonant peaks which diminish as the particle number is increased. Moreover the distribution decays much quicker as $W\to 0$ when interactions are present. 
 By expanding the Bessel functions and using the identity 
$\sum_{P}(-1)^P(\|P\|^2)^n=0,~\forall n< \binom{N}{2}$\cite{Speer}, we have that near the threshold when the system is interacting
\begin{eqnarray}\label{Edge}
\mathcal{P}(W)\sim W^{\frac{N^2}{2}-1}.
\end{eqnarray}
This exponent can be interpreted as coming from the coalescence of all the edge singularities in the gapped case. As with the gapped case this dramatically different when there are no interactions present. Repeating  the same calculation in the non interacting case gives instead that $\mathcal{P}(W)\sim W^{\frac{N}{2}-1}$. The value of the exponent is not dependent on the value of the interaction strength other than if it is non zero. This highlights how even weakly interacting theories exhibit strong correlations in one dimension.

The region $W\sim 0$ of the distribution gives insight to the long time behavior of the Loschmidt echo. In the presence of interactions we have that as $t\to \infty$, $|\mathcal{G}(t)|^2\to1/t^{N^2}$ whilst in the non interacting case we have $|\mathcal{G}(t)|^2\to1/t^{N}$. The much faster decay in the interacting case results from the process of dynamical fermionisation\cite{IyerGuan, Iyer} wherein the interacting bosons acquire fermionic correlations as the system evolves. Consequently the bosons spread out through the trap and result in a rapidly vanishing overlap with  $\ket{\Psi_i}$.



\subsection{Attractive interactions}
We now turn to the attractive regime which is of significant interest.  The properties of the attractive model both in and out of equilibrium are much less studied than its repulsive counterpart.
This dearth of theoretical results stems from the increased complexity of the Bethe Ansatz solution in the attractive model.  When $c<0$ the model supports bound states and the ground state consists of a single bound state of all $N$ particles \cite{Maguire}. While the eigenstates given by  \eqref{states} and Bethe equations  \eqref{BAE} remain valid, complex values of $k$ which correspond to bound states are allowed. The resolutions of the identity appearing in  \eqref{Identity} and \eqref{IdentityY} also remain formally valid provided these complex valued solutions are accounted for \cite{PiroliCalabreseEssler, Zill}. A large stumbling block however is that the normalisation of the Bethe states in the attractive regime is not known in closed form. 

In the low density limit however it has been shown that for both repulsive and attractive interactions the spatially ordered identity \eqref{IdentityY} becomes
 \cite{Yudson,Iyer, IyerGuan, AndreiQuench}
\begin{eqnarray}\label{IdentityIyer}
\mathbb{1}_N=\int_\Gamma\frac{\mathrm{d}^Nk}{(2\pi)^N}\ket{\{k\}}\bray{\{k\}}
\end{eqnarray}
where we label the eigenstates by $\{k\}$ rather than $\{n\}$. The contours of integration, $\Gamma$ lie on the real line for repulsive interactions and are spread out in the imaginary direction for the attractive case with Im$(k_{j+1}-k_{j})>|c|$. 

Making use of this here in conjunction  with the same $|c|\gg m\omega$ expansion we find that  the work done in the attractive regime separates into two contributions,
\begin{eqnarray}\label{Patt}
\mathcal{P}_{c<0}(W)&=&\mathcal{P}_{\text{free}}(W)+\mathcal{P}_\text{bound}(W).
\end{eqnarray}
 The first term $\mathcal{P}_{\text{free}}(W)$ is the contribution from particles which do not form bound states, 
it is identical to the expression in repulsive case given in \eqref{Prep} only now $c<0$. The major difference imposed by this is that the effective distance between the particles is smaller $\delta_\text{eff}<\delta$, the attractive interactions promoting the clustering of particles.  

The  simple analytic continuation to negative coupling of the first term is  reminiscent of the the super Tonks-Girardeau gas\cite{SuperTG1,SuperTonks2, SuperTGExp}. This highly  correlated state of the LL model is created by preparing a repulsive LL gas in the Tonks-Girardeau limit, $c\to\infty$ \cite{Tonks, Girardeau} and then  abruptly changing the interaction strength from the being large and positive to large and negative. The result is a metastable nonequilibrium state which exhibits enhanced correlations. Many of the properties of this state can be shown to emerge from a simple analytic continuation of the coupling to large negative values \cite{SuperTG3,SuperTG4, Muth}.
In effect the negligible overlap of each particle of our initial state mimics the density profile of the TG gas and so super-TG like behaviour is not unexpected. We should stress  that the expression \eqref{IdentityIyer} is valid at arbitrary negative values $c$ and so not limited to super-TG regime.

The second term  $\mathcal{P}_\text{bound}(W)$ is entirely different. It is due to the bound states and is  calculated by deforming the contours in \eqref{IdentityIyer} to the real line and picking up  contributions due to the poles at $k_i-k_j=i c$ present in in \eqref{states}.  
An $n$-particle bound state can be shown to contribute $\mathcal{P}_{n-\text{bound}}(W)\propto |c|^{n-1}e^{-n|c|\delta}$ with factors from multiple bound states being multiplicative. 

This exponential factor means that the probability that the initial state transitions to one containing bound states is highly suppressed and in the true super-TG limit vanish entirely.  
Despite this, for finite $|c|$ the bound states have a strong signature in work distribution function. Since forming a bound state will lower the energy of the system\cite{Maguire} the work distribution becomes non vanishing at negative values of $W$. There is a non zero probability that work can be extracted from the system. Importantly this does not violate the 2nd law of thermodynamics as the average work remains positive $\left<W\right>$ \cite{Jarzynski1, Jarzynski2}.  In fact, it has been shown observed recently that the probability of extracting work from a single electron transistor can be as high as 65\% whilst still satisfying the 2nd law \cite{Maillet}.  

To see this we examine the leading term of $\mathcal{P}_\text{bound}(W)$ which arises due to the formation of a single two particle bound state 
 \begin{eqnarray}\label{Pbound}
   \mathcal{P}_\text{bound}(W)\approx N\sqrt{\frac{2\pi\omega}{m}}\frac{e^{-|c|\delta-\frac{2W}{\omega}}}{\Gamma\left(\frac{N}{2}-1\right)}\left[\frac{2(W+\frac{|c|^2}{4m})}{\omega}\right]^{\frac{N}{2}-2}
.
 \end{eqnarray}
Which is non vanishing for $-|c|^2/4m<W$. Determining the full bound state contribution is a straightforward yet involved calculation which we we will not deal with here.

\section{Finite Density and the role of boundary conditions}

It is interesting to study also the case where $\rho=\delta$ so that the initial state consists of a fully occupied lattice. In this scenario the boundary conditions play an important role and will change the behavior of the system in the region $W\ll \left<W\right>$. At finite density the Bethe equations can be used to reduce many of the terms in the sum over permutations in $G(\{n\})$ \eqref{Gexact}. For example the permutation correspond to the ordering $P=(23\dots N1)$ gives 
\begin{eqnarray}\label{GPBC}
e^{-i(N-1)\delta k_1-i\sum_{j}\varphi(k_1-k_j)}&=&e^{i\delta\sum_jk_j}
\end{eqnarray}
where we have used $L=N\delta$ and \eqref{BAE} to get the right hand side. Taking the same limits as before one arrives at \eqref{GQ} for the Loschmidt amplitude however the boundary conditions should be taken into account when calculating $\alpha_P^2$. For instance when $\rho\ll\delta$ the permutation $P=(23\dots N1)$ gives $\alpha_P^2=m\delta^2_\text{eff}N(N-1)/2$ however for periodic boundary conditions as a result of \eqref{GPBC} we get instead $\alpha_P^2=m\delta^2N/2$. The terms corresponding to no or few particles exchanging positions are unaffected by the boundary conditions however they serve to reduce those which involve widely separated particles exchanging positions. 

The work distribution function is again given by \eqref{Prep} in this case however as with the Loschmidt amplitude one must account for the boundary conditions when calculating $\alpha_P^2$. In the region of $W\sim \left<W\right>$ this has little effect as this area is dominated by terms corresponding to no or few exchanges of particles and the average work is as before. At small values of work however all permutations contribute and there is a difference. By expanding the Bessel functions to first order  we find that instead
\begin{eqnarray}
\mathcal{P}(W)\sim W^{\frac{N}{2}}.
\end{eqnarray} Consequently the long time decay of the echo is given by $|\mathcal{G}(t)|^2\to 1/t^{N+2}$. In this instance the strongly interacting particles have no large trap to expand into as was the case before resulting in slower decay of the echo. 

\section{Conclusion}
In this paper we have studied the work done on an interacting cold atomic gas during a quantum quench. Specifically we studied the work done on the system when it is fully released form an optical lattice. Using Bethe Ansatz methods we derived an exact expression for the Loschmidt amplitude valid at all values of the interaction strength and using this studied  the work distribution function at large $|c|\gg m\omega$. The edge exponents and average work were calculated and we showed that when the gas is released in a much larger trap the edge exponents differ dramatically when the system is interacting. As a consequence the Loschmidt echo decays much more rapidly in the interacting system.  This calculation was then extended to the attractive regime where it was seen that the formation of bound state in the post quench system result in a small, non vanishing probability of extracting work form the optical lattice quench. Furthermore the distribution displayed properties which are indicative of the super Tonks-Girardeau gas, meaning that the non bound state contribution is obtained from analytically continuing the repulsive result to negative coupling. Finally we examined the case where the initial lattice has unit filling and periodic boundary conditions. Here it was seen that although the edge exponent did change when interactions were present the effect was not as dramatic as the case of open boundary conditions. 

The calculations presented here were carried out for the simplest case of at most one particle per lattice site in the initial state. The results, especially in the attractive regime suggest that starting from higher filling number or using a model with more bound state channels such as the Gaudin-Yang gas would produce interesting results\cite{Yang1, Gaudin, GuanAndrei}. 

The experimental measurement of the work statistics of a closed quantum system after a quench has been proposed and carried out in a number of different settings \cite{Tureci, Tureci2, Goold2, Maillet} including in cold atom gases\cite{QuantumWorkmeter}. Although this has so far not been carried out in the setting we have proposed our results provide firm predictions for what would be observed.
\acknowledgements{}
This research was supported by DOE-BES (DESC0001911) (CR) an by NSF Grant DMR 1410583 (NA).
\bibliography{mybib}
\end{document}